%% file: hoi_pra_20210416.tex
\documentclass[twocolumn,superscriptaddress,aps,showpacs,floatfix,pra,10pt]{revtex4-1}

\usepackage[T1]{fontenc}
\usepackage{amssymb,amsmath}
\usepackage{graphicx}
\usepackage{color}
\usepackage{bbm}
\usepackage{bbold}
\usepackage{soul}
\usepackage[dvipsnames]{xcolor}
\usepackage{ulem}
\begin{document}

\title{Higher-order interference { between multiple quantum particles interacting nonlinearly}}

\author{Lee A. Rozema}
\affiliation{Vienna Center for Quantum Science and Technology (VCQ), Faculty of Physics, University of Vienna, Boltzmanngasse 5, Vienna A-1090, Austria}

\author{Zhao Zhuo} 
\affiliation{School of Physical and Mathematical Sciences, Nanyang Technological University, Singapore 637371, Singapore}

\author{Tomasz Paterek}
\affiliation{School of Physical and Mathematical Sciences, Nanyang Technological University, Singapore 637371, Singapore}
\affiliation{MajuLab, International Joint Research Unit UMI 3654, CNRS, Universite Cote d'Azur, Sorbonne Universite, National University of Singapore, Nanyang Technological University, Singapore}
\affiliation{Institute of Theoretical Physics and Astrophysics, Faculty of Mathematics, Physics and Informatics, University of Gda\'nsk, 80-308 Gda\'nsk, Poland}

\author{Borivoje Daki\'c}
\affiliation{Vienna Center for Quantum Science and Technology (VCQ), Faculty of Physics, University of Vienna, Boltzmanngasse 5, Vienna A-1090, Austria}
\affiliation{Institute for Quantum Optics and Quantum Information (IQOQI), Austrian Academy of Sciences, Boltzmanngasse 3, A-1090 Vienna, Austria}

\begin{abstract}
The double-slit experiment is the most direct demonstration of interference between individual quantum objects.
Since similar experiments with single particles and more slits produce interference fringes reducible to a combination of double-slit patterns it is usually argued that quantum interference occurs between pairs of trajectories, compactly denoted as second-order interference.
Here we show that quantum mechanics in fact allows for interference of arbitrarily high order.
This occurs naturally when one considers multiple quantum objects interacting in the presence of a nonlinearity, both of which are required to observe higher-order interference.
We make this clear by treating a generalised multi-slit interferometer using second-quantisation. 
We then present explicit experimentally-relevant examples both with photons interacting in nonlinear media and an interfering Bose-Einstein condensate with particle-particle interactions.
These examples are all perfectly described by quantum theory, and yet exhibit higher-order interference {based on multiple particles interacting nonlinearly}.
\end{abstract}

\maketitle

Quantum states are represented by density matrices, whose elements can be estimated in a series of interference experiments involving a superposition of only two basis states at a time.
Already at this abstract level one expects that any interference pattern should be fundamentally reducible to two-state interference.
Indeed, it was shown theoretically, within the framework of generalised measure theories, that interference fringes observed in multi-slit experiments are simple combinations of patterns observed in double-slit and single-slit experiments.
Quantum mechanics has hence been termed a ``second-order interference theory''~\cite{Sorkin}.
It is possible, however, to devise a family of post-quantum theories exhibiting higher-order interference {\cite{zyczkowski2008quartic,dakic2014density,lee2017higher}}. Motivated by this, experiments based on photons~\cite{Weihs2010,Weihs2011,Weihs2017,hickmann2011}, nuclear magnetic resonance~\cite{Laflamme2012}, spins in diamond NV centers \cite{jin2017} and with matter waves~\cite{Arndt2017,Barnea2018} have placed bounds on higher-order interference, verifying, within experimental error, that in these setups higher-order interference is absent.
Atomic analogs of multi-slit experiments have also been studied in detail~\cite{Lee2019}.
{The presence of such higher-order interference is often discussed in the literature as violation of Born's rule \cite{Weihs2010}. Nevertheless, whether viewed as a violation of Born's rule or as higher-order interference, such a finding would require an explanation that goes beyond our current formulation of quantum mechanics.}

While the original theoretical work showing that quantum mechanics is a second-order theory considered a single electron incident on a multi-slit, many experimental tests have used multi-particle states. For example, in photonic experiments both single photons and multi-partite coherent states have been used~\cite{Weihs2010,Weihs2011,Weihs2017}, Ref.~\cite{jin2017} used ensembles of spin-1/2 particles in an NMR experiment, while the experiments presented in Refs.~\cite{Arndt2017,Barnea2018} used thermal states of atoms and molecules.
Therefore, although use of single-particle states was implicit in the original theory, the effect of multi-partite states was not appreciated at the experimental level.

{
Very recent work has focused on this problem, and derived new limits on higher-order interference (via the application of Born's rule) when the input state is a multiple-particle state with a fixed number of particles~\cite{Pleinert2020theor}.
In these results, new quantities and measurements based on multi-particle input states were derived~\cite{Pleinert2020theor} and experimentally probed~\cite{Pleinert2020expt}, allowing for more sensitive tests of higher-order interference and Born's rule.
}

{
Here we take a slightly different approach, more similar to that of the recent so-called looped-trajectories results \cite{Yabuki86,RMH2012,Sinha2014,Sinha2015}.
In these works, a multi-slit apparatus and the ``standard'' measurements used to search for higher-order interference are considered.
It was then shown that the multi-mode character of an actual slit experiment can lead a small third-order interference term~\cite{Yabuki86,RMH2012,Sinha2014,Sinha2015}.
The origin of this term lies in how the superposition principle is applied, and rests on different boundary conditions in multi-slit and single-slit setups.
This effect has even been experimentally confirmed~\cite{Boyd2016,Sinha2018}.
Similarly, here we show that given a multi-slit apparatus and the standard measurements, multi-partite states and a nonlinear interaction (such as an optical nonlinearity or particle-particle interaction) can lead to the emergence of apparent higher-order interference.
However, the third-order interference term both in our work, and the looped-trajectory work is not due to ``post-quantum higher-order interference'', but rather comes from implicit assumptions in the theory that are not met experimentally.
Hence, if one wishes to bound the contribution of genuine higher-order interference present in a given experiment, all such effects must be considered.}

{
We also briefly show below, that the experimental arrangement of Refs. \cite{Pleinert2020theor,Pleinert2020expt} can lead to apparent higher-order interference and the underlying reason is nonlinearity. 
It is the nonlinearity in the detection, which looks for coincidences of various detection events and is hence by construction nonlinear in the incident photon number.
}

{ Throughout our paper we will refer to the interference studied here as higher-order interference. We use this terminology because these effects could arise in any experiment used to search for genuine higher-order interference, given sufficiently strong nonlinearities and multi particle states. 
We stress that our higher-order interference does not emerge if just multi-partite (or multi-photon) states are used or just in the presence of nonlinear elements; rather, both features are required and the apparatus must be used to construct the so-called M-path interference introduced in Eq.~(\ref{EQ_I_M}).
We show this can lead to very strong observable deviations from vanishing higher-order interference, even when the issues due to boundary conditions are negligible.}
As is the case for the looped trajectories, we emphasise that it is also not genuine post-quantum higher-order interference, but rather an artifact of multi-particle interactions.
Our results stress the need for high-quality single-particle sources in experiments searching for such deviations from quantum theory.

Our paper is organised as follows.
We will first {use our formalism to} show that all single-particle states give rise to only second-order interference, {as was already shown in} Ref.~\cite{Sorkin}.
We will then introduce a means to quantify the order of interference in the framework of second quantization, and then  show that all linear processes are limited to second-order interference.
Finally, we will provide explicit examples of nonlinear processes with multi-particle input states that produce interference of arbitrary order.
The required nonlinearity can be caused by different physical mechanisms, and we show examples of higher-order interference based on optical nonlinearity, nonlinear detector response, and particle-particle interaction in a Bose-Einstein condensate modeled by the Gross-Pitaevskii equation.


\begin{figure}[!t]
\centering
\includegraphics[width=.6\linewidth]{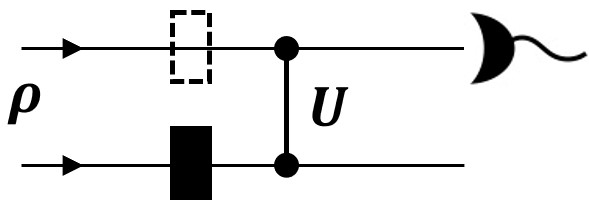}
\caption{Measuring interference. 
The simplest interference experiment involves two paths that can be individually blocked.
This figure shows a configuration where the upper path is open and the lower path is blocked.
Here $\rho$ denotes the input state, and \textbf{U} is some unitary interaction between the two modes.
Interference is present if the intensity measured with both paths open is different from the sum of the intensities measured with the individual paths open.}
\label{FIG_2PATHS}
\end{figure}

\section{The order of interference} 

Consider first an experiment with only two paths, see Fig.~\ref{FIG_2PATHS}.
Each path can either be open (0) or blocked (1), so that the configuration in Fig.~\ref{FIG_2PATHS} is represented by sequence 01.
Interference is said to occur if the mean number of particles (the intensity) measured with both paths open, $I_{00}$, is different from the sum of the intensities for individual paths blocked, $I_{01} + I_{10}$.
It is hence natural to quantify two-path interference by $\mathcal{I}_2 \equiv I_{00} - I_{01} - I_{10} + I_{11}$, where we have introduced $I_{11} = 0$ for symmetry reasons.
A similar argument applied to a scenario with $M$ paths leads to the definition of $M$-path interference when the following quantity is non-zero~\cite{Sorkin}:
\begin{equation}
\mathcal{I}_M = \sum_{x_1, \dots, x_M = 0}^1 (-1)^{x_1 + \dots + x_M} I_{x_1 \dots x_M}.
\label{EQ_I_M}
\end{equation}
Sorkin showed that when $\mathcal{I}_M = 0$ for some $M$, then all the quantities $\mathcal{I}_M$ with higher $M$ also vanish~\cite{Sorkin}.
The highest index $M$ for which a theory predicts non-zero $\mathcal{I}_M$ is then called the order of interference of that theory.

Classical particle experiments do not give rise to any form of interference; i.e. classically, one already has $\mathcal{I}_2 = 0$.
In the quantum case, consider first $N$ particles sent one-by-one into the setup in Fig.~\ref{FIG_2PATHS}.
Each particle is in state $\rho$ spanned by kets $| u \rangle$ and $| d \rangle$, describing propagation along upper and lower path respectively.
The intensity $I_{00}$ is given by $N p_{00}$, where $p_{00}$ is the probability that the particle is in the upper path after the unitary, i.e. $I_{00} = N \langle u | U \rho U^\dagger | u \rangle$.
Similarly $I_{01} = N \rho_{uu} \langle u | U | u \rangle \langle u | U^\dagger | u \rangle$ and 
$I_{10} = N \rho_{dd} \langle u | U | d \rangle \langle d | U^\dagger | u \rangle$, where e.g. $\rho_{uu} = \langle u | \rho | u \rangle$.
One finds that the second-order interference, $\mathcal{I}_2$, vanishes for all (classical) states that exhibit no coherence, $\rho_{ud} = 0$, independent of the choice of the unitary $U$.

On the other hand, particles in quantum states can undergo second-order interference.
The maximum second-order interference is $\mathcal{I}_2 = N/2$, which is achieved for the input state $(| u \rangle + | d \rangle) / \sqrt{2}$ and the unitary describing a 50-50 beam-splitter, as expected.
Applying the same calculation to three-path experiment (with an arbitrary unitary acting on all three paths) shows that $\mathcal{I}_3 = 0$ for all input states and all unitaries.
This leads to the usual statement that quantum theory is a second-order interference theory.
However, we will now show that multiple quantum systems interacting nonlinearly can lead to non-zero $\mathcal{I}_M$ for arbitrary $M$.


\section{Interference of indistinguishable particles} 

We first show that all linear processes give rise to only second-order interference, i.e. $\mathcal{I}_3 = 0$, independently of the input multipartite state.
{Although this is already evident from Ref.~\cite{Sorkin}, we will now show it usung our formalism.}
A linear process is described by $U=e^{iH}$, where $H$ is linear in the ladder operators: $H=\sum_{n,m} h_{nm}a_n^{\dagger}a_m$ \cite{gerry2005introductory}.
We consider a beam of indistinguishable particles 
in any state $\rho$ loaded into a setup with $M$ paths; later we will specify $M=3$.
The particle number does not have to be well defined, e.g. the input could be a series of coherent states of photons in the various input modes.
With each path we associate local Fock space $\mathcal{H}_m$ spanned by Fock states $| n \rangle_m$ describing $n$ excitations (photons) in the $m$th path (mode).
The entire Hilbert space of this system is therefore a tensor product $\mathcal{H}_1 \otimes \dots \otimes \mathcal{H}_M$ and we need to specify how blocking paths is represented in this formalism.
Appendix~\ref{APP_A} shows that the operation of blocking the $m$th path, $\Pi_m$, has the following intuitive effect: $\Pi_m(\rho) = | 0 \rangle_m \langle 0 | \otimes \mathrm{Tr}_m(\rho)$,
i.e. it produces vacuum in the blocked path and decorrelates it from all the other paths 
(here $\mathrm{Tr}_m(\rho)$ stands for the partial trace over the Fock space $\mathcal{H}_m$, meaning that the states in the other paths are unaffected by the blocker).
With this notation the state after the blockers is given by
\begin{equation}
\rho_{x_1 \dots x_M} = \Pi_1^{x_1} \otimes \dots \otimes \Pi_M^{x_M} (\rho),
\end{equation}
where we set $\Pi_m^0$ to the identity operator. Recall $x_i$ represents whether the blocker in mode $i$ is present or not.
By placing the final detector along the first path, the intensities can be computed from
\begin{equation}
I_{x_1 \dots x_M} = \mathrm{Tr}(a_1^\dagger a_1 U \rho_{x_1 \dots x_M} U^\dagger),
\end{equation}
where $a_1^\dagger a_1$ is the number operator in $\mathcal{H}_1$.
Since the same measurement is conducted for all the combinations of blocked paths,
we introduce an ``interference operator'' via the relation $\mathcal{I}_M = \mathrm{Tr}(U^{\dagger} a_1^{\dagger} a_1 U \hat{\mathcal{I}}_M )$, see Appendix~\ref{APP_B} for its explicit form and properties.
Any linear process satisfies $U^\dagger a_1^\dagger U = \sum_m u_{m} a_m^\dagger$ and accordingly
\begin{equation}
\mathcal{I}_3 = \sum_{m,m'} u_{m} u_{m'}^* \mathrm{Tr}(a_m^\dagger a_{m'} \hat{\mathcal{I}}_3 ) = 0,
\label{EQ_I30}
\end{equation}
where in the last equation we used the fact that the interference operator vanishes under the partial trace over any path, see Appendix~\ref{APP_B}.
Hence, we see that if the interaction is linear or the input state is a single-particle state we have $\mathcal{I}_3=0$.

In general, the same line of reasoning applies to nonlinear processes and higher-order interference terms.
A process of order $k$, i.e. where the creation operator $a_1^\dagger$ is mapped to a polynomial $\sum_{m_1,\dots, m_k} u_{m_1 \dots m_k} a_{m_1}^\dagger \dots a_{m_k}^\dagger$,
gives rise to vanishing higher-order interference terms $\mathcal{I}_M$ with $M > 2k$.
Note that non-linearity is necessary for higher-order interference, but not sufficient.
For example, a non-linear process mapping $a_1^\dagger$ to a sum of squared operators $\sum_m u_{m} a_m^\dagger a_m^\dagger$ still admits $\mathcal{I}_3 = 0$,
because in Eq. (\ref{EQ_I30}) each term in the sums couples only two paths and hence the partial trace argument gives vanishing $\mathcal{I}_3$.
Thus, experimentally finding a non-zero $\mathcal{I}_M$ indicates the presence of nonlinear multi-mode coupling in the underlying process (which could be completely unknown, i.e. a black box) and provides the minimal number of the coupled paths.

\begin{figure}[!t]
\centering
\includegraphics[width=.9\linewidth]{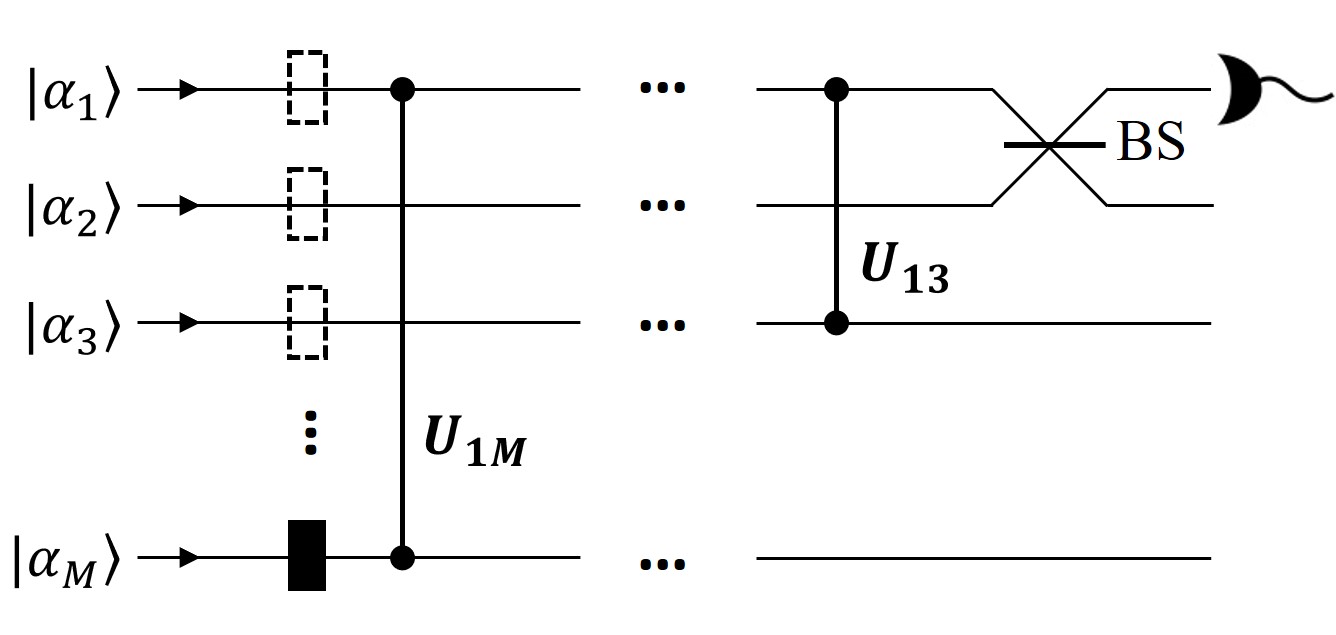}
\caption{Schematic for measurement of higher-order interference in quantum optics.
The input coherent states propagate through blockers and then through a sequence of nonlinear phase shifters $U_{1M}, \dots, U_{13}$.
The final element, marked BS, is a 50-50 beam splitter between the first and the second path, and the intensity is monitored in the first path.
This setup gives rise to the $M$th-order interference, see Eq. (\ref{EQ_MTH}).}
\label{FIG_M}
\end{figure}

In the same way one recovers the recent results of Ref.~\cite{Pleinert2020theor} on multi-partite higher-order interference.
In the present notation, their $n$-partite $M$-th order interference term is $\mathcal{I}_M^n = \mathrm{Tr} (a_1^\dagger \dots a_n^\dagger a_n \dots a_1 U \mathcal{\hat I}_M U^\dagger)$, with the interference operator as introduced above.
Therefore, again due to the partial trace argument, $\mathcal{I}_M^n = 0$ for $M > 2 n$ when $U$ is a linear process.
Interestingly, it is also apparent that when $U$ is nonlinear, quantum theory predicts that the multi-partite higher-order interference, defined in Ref.~\cite{Pleinert2020theor}, may be non-zero.

\section{Nonlinear phase shift}

Now we give an example of a nonlinear process from quantum optics (a two-mode nonlinear phase shifter, i.e. cross-phase modulation) whose concatenation gives rise to arbitrarily high order of interference.
The exact setup is presented in Fig.~\ref{FIG_M}.
The unitary describing this nonlinear process between modes $j$ and $k$ has the following effect:
\begin{eqnarray}
U_{jk}^\dagger a_j^\dagger U_{jk} & = & a_j ^\dagger \exp(- i \theta a_k ^\dagger a_k), \\
U_{jk}^\dagger a_k^\dagger U_{jk} & = & a_k ^\dagger \exp(- i \theta a_j ^\dagger a_j),
\end{eqnarray}
where $\theta$ is the strength of non-linearity.
{ It can range from $\approx 10^{-18}$ rad/photon for a bulk Kerr media \cite{boyd1999order,barrett2005symmetry}, to $\approx 10^{-7}$ rad/photon using a photonic crystal fibres \cite{matsuda2009observation}, all the way up to $\approx 10^{-2}$ rad/photon using electromagnetically-induced transparency, e.g.~\cite{EIT1,EIT2}.}
For simplicity, we will assume that all the input coherent states have the same mean number of photons, $\langle n \rangle = |\alpha|^2$, but potentially different phases. In this case, the setup in Fig.~\ref{FIG_M} produces the following value of higher-order interference (see Appendix~\ref{APP_C} for details):
\begin{equation}
\mathcal{I}_M = \left| \langle n \rangle \left( \exp [- \langle n \rangle (1 - e^{- i \theta})] - 1\right)^{M-2} \right| \cos(\varphi_2 - \varphi_1 - \delta),
\label{EQ_MTH}
\end{equation}
where $\varphi_1$ and $\varphi_2$ are the phases of the input light along the first two paths and $\delta$ is a fringe offset. 
The phases of the remaining inputs do not enter the interference formula.
{ 
For example, to achieve $\mathcal{I}_3 \sim 1$ with natural Kerr nonlinearity of $10^{-18}$ one requires a mean photon number of about $10^{9}$ photons per mode. If one uses a pulsed laser system this corresponds to a pulse energy of $\approx 0.2$ nJ at a wavelength of $1000$ nm, which is easily available in commercial laser systems.}

At this stage we would like to comment on two recent experiments which may seem related.
Refs.~\cite{Jennewein2017,Walmsley2017} observed genuine three-photon interference as a generalisation of the famous Hong-Ou-Mandel dip~\cite{HOM}.
{Although these works used (nonlinear) multi-photon detection and multi-partite input states}, the higher-order interference we describe here is distinct from the observed multi-photon interference.
{ Indeed, when the setup in Fig.~\ref{FIG_M}, restricted to $M=3$ paths, is input with the field in the state $\frac{1}{\sqrt{2}}(|01\rangle_{12} + |10\rangle_{12}) \otimes \frac{1}{\sqrt{2}}(|0 \rangle_3 + |1 \rangle_3)$
the third-order interference term reads $\mathcal{I}_3 =  - \sin^2 (\theta/2)$.
While the number of photons is not fixed in this state, it is either $1$ or $2$ in each branch of the input superposition and therefore the third-order interference could be observed with vanishing probability of detecting more than two photons.}

Before presenting examples of other nonlinearities that produce higher-order interference, let us note that due to the specific combination of terms entering the higher-order interference expression, any noise that is independent of the input signal is irrelevant, e.g. detector dark counts.
If the noise alone is characterised by probability $d(n)$ to observe $n$ photons and the ideal signal has probabilities $p(n)$, the independence is encoded by the convolution
$r(n) = \sum_{k} p(k) d(n-k)$, where $r(n)$ is the probability of observing $n$ photons with the noisy detector.
Such noise just shifts the intensity of arbitrary input state by the same amount $\Delta$, see Appendix~\ref{APP_D}.
Therefore, the higher-order interference term in the presence of noise is given by
\begin{equation}
\tilde{\mathcal{I}}_M = \sum_{x_1, \dots, x_M = 0}^1 (-1)^{x_1 + \dots + x_M} (\Delta + I_{x_1 \dots x_M}) = \mathcal{I}_M.
\end{equation}
This shows the robustness of estimating higher-order interference in a real laboratory setting.

\begin{figure}[!t]
\centering
\includegraphics[width=.9\linewidth]{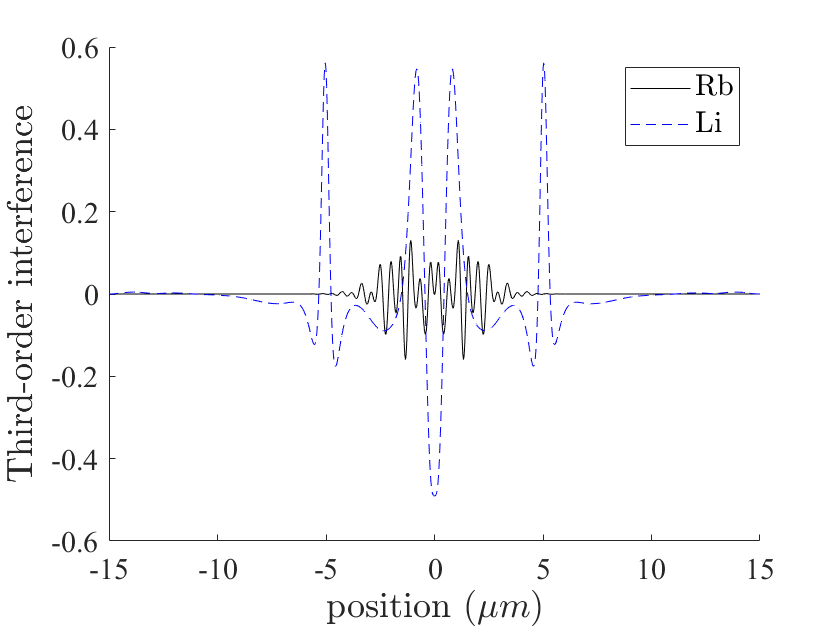}
\caption{Higher-order interference with Bose-Einstein condensates.
The condensate is initially prepared in an even superposition of gaussian wave functions each with {position spread (standard deviation) of} $1 \mu$m centered at $\pm 5 \mu$m and $0$. 
The solid line presents the values of $\mathcal{I}_3$ for the repulsive condensate of $^{87}$Rb with parameters $N = 10^3$ atoms, $a = 5.8$ nm and $m = 1.45 \times 10^{-25}$ kg after an evolution time of $\tau=1$~ms.
The dashed line is for the attractive condensate of $^{7}$Li with parameters $N = 500$ atoms, $a = -1.2$ nm and $m = 1.16 \times 10^{-26}$ kg.
Both are experimentally feasible with present day technology~\cite{Dalibard1999,Brachet1999}.
}
\label{FIG_GPE}
\end{figure}

\section{Interacting Bose gas} 

Nonlinearity in other physical systems can also lead to higher-order interference.
For example, consider a Bose-Einstein condensate initialised in an even superposition of three Gaussian wave functions.
We compute its one-dimensional dynamics according to the nonlinear Gross-Pitaevskii equation
\begin{equation}
i \hbar \frac{\partial \psi(x,t)}{\partial t} = -\frac{\hbar^2}{2m} \frac{\partial^2 \psi(x,t)}{\partial x^2} + \frac{4 \pi \hbar^2 N a}{m} |\psi(x,t)|^2 \psi(x,t),
\end{equation}
where $N$ is the number of atoms, each of mass $m$, and $a$ is the scattering length.
The initial wave function is normalised to unity.
The system is evolved for a time $\tau$ after which we record the distribution of particles in one dimension.
Blocking the paths is modeled by removing the corresponding part of the initial superposition (and keeping the state unnormalised).
Fig.~\ref{FIG_GPE} shows the results for $\mathcal{I}_3$ confirming the experimental feasibility of observing third-order interference. 
The same conclusion is expected to hold in other physical systems with dynamics modeled by the Gross-Pitaevskii equation, e.g. polaritons~\cite{RMP.85.299}.

\emph{Detection nonlinearity.} Our last example is a photodetector with a nonlinear response.
The main features of a real detector are: an essentially linear response in the low intensity regime and saturation, which may set in for high input intensities.
To illustrate our point, we are only interested in the saturation domain where the measured intensity $I_r$ is modeled as $I_r = I_i - \epsilon I_i^2$, where $I_i$ is the output of an ideal detector and $\epsilon$ is the strength of nonlinear saturation.
For single-photon detectors $\epsilon$ is given by the detector dead time; this can be seen by expanding equation (A1) of \cite{Kauten2014}.
With such detectors, even linear interactions can give rise to non-zero higher order interference; this has been discussed from an experimental perspective in Refs.~\cite{Kauten2014,Weihs2017}.
We now provide a simple theoretical example in which a non-zero $\mathcal{I}_3$ appears in a setup with three paths combined on a symmetric tritter (a generalisation of a beam splitter to three paths).
In particular the unitary describing the tritter has matrix elements given by $U_{ij}=\frac{1}{\sqrt d}\omega^{ij}$, where $\omega=e^{i2\pi/3}$.
After the tritter we monitor the first output port with the nonlinear detector.
Accordingly, the tritter unitary gives $U^\dagger a_1^\dagger U = \frac{1}{\sqrt{3}}(a_1^\dagger + a_2^\dagger + a_3^\dagger)$ and the detector is represented by $a_1^\dagger a_1 - \epsilon (a_1^\dagger a_1)^2$. With this at hand one finds a third-order interference term of $\mathcal{I}_3 = - 4 \epsilon |\alpha|^4$, where we have also assumed that all three input modes are injected with the same coherent states $| \alpha \rangle$.
Taking a dead time of $50$ ns (which is a typical value for commercially available single-photon detectors), this leads to $|\alpha|^2\approx 2000$ for $\mathcal{I}_3 = 1$, which can be understood as the number of photons per detector deadtime. This is equivalent to $\approx 10^{10}$ photons per second or about $10$ nW.

\section{Conclusions}

We have theoretically demonstrated the emergence of higher-order interference within the standard formalism of second quantisation. Its origin is traced to nonlinearity in multipartite processes.
However, if the interaction is linear or the input state is a single-particle state then $\mathcal{I}_3=0$.
Moreover, the non-vanishing $\mathcal{I}_3$ should be observable with present day technology such as with nonlinear optics or Bose-Einstein condensates.
Our work shows that if one wishes to place limits on quantum theory, nonlinearities elsewhere in the system must be considered, and single-particle states should be used in the experiments.

Finally, it is worth stressing the difference between our higher-order interference and that arising from looped trajectories~\cite{Yabuki86,RMH2012,Sinha2014,Sinha2015,Boyd2016,Sinha2018}.
Looped trajectories arise as a consequence of real-world multi-slit experiments being multi-mode devices.  
In fact, it has been pointed out that if one replaces the traditional triple-slit experiment with single-mode beams interfering on a tritter (or some other unitary structure) the higher-order interference due to looped trajectories becomes negligible~\cite{Weihs2017}.
This is exactly the case in our proposal, which deals with $M$ ideal single modes which do not admit such exotic trajectories.
Hence, the high-order interference that we predict cannot be understood as a systematic experimental error but is fundamental to multipartite nonlinear quantum systems.
Nevertheless, both our nonlinear examples and the looped trajectories illustrate that different implicit assumptions are made in the claim that quantum mechanics is a second-order interference theory, and have direct consequences for experiments searching for higher-order interference.

\section*{Acknowledgments}
We thank M. Radonji\'c and \v C. Brukner for useful discussions.
L.A.R. acknowledges support from the Templeton World Charity Foundation (fellowship no. TWCF0194) and the Austrian Science Fund (FWF) through BeyondC (F71).
T.P. is supported by the Polish National Agency for Academic Exchange NAWA Project No. PPN/PPO/2018/1/00007/U/00001.
B.D. acknowledges support from an ESQ Discovery Grant
of the Austrian Academy of Sciences (OAW) and the Austrian Science Fund (FWF) through BeyondC (F71).

\section*{Appendix}

\appendix

\subsection{Path blocker in second quantisation}
\label{APP_A}

We prove that $\Pi_1(\rho) = | 0 \rangle_1 \langle 0 | \otimes \rho_{2\dots M}$, where $\rho_{2\dots M}$ is the reduced state on all the other paths.
First of all, clearly $\Pi_1(\rho_1) = | 0 \rangle_1 \langle 0 |$, i.e. any state injected to the blocked path results in the vacuum on that path.
Arbitrary state $\rho$ can be decomposed as $\rho = \sum_{j,k} c_{jk} \rho_1^{(j)} \otimes \rho_{2 \dots M}^{(k)}$, where the coefficients are not necessarily non-negative, but all the summed matrices are proper quantum states.
Since $\Pi_1$ is a linear operation, we have $\Pi_1(\rho) = | 0 \rangle_1 \langle 0 | \otimes \rho_{2 \dots M}$ as claimed.

\subsection{Properties of the interference operator}
\label{APP_B}

The interference operator is defined via relation $\mathcal{I}_M = \mathrm{Tr}(U^{\dagger} a_1^{\dagger} a_1 U \hat{\mathcal{I}}_M )$.
It therefore has the following explicit expansion
\begin{equation}
\hat{\mathcal{I}}_M = \sum_{x_1, \dots, x_M = 0}^1 (-1)^{x_1 + \dots + x_M} \rho_{x_1 \dots x_M}.
\end{equation}
Its crucial property used in our arguments is
\begin{equation}
\mathrm{Tr}_k(\hat{\mathcal{I}}_M) = 0,
\label{EQ_PROP}
\end{equation}
where the partial trace is over arbitrary subset of paths denoted collectively as $k$.

\subsection{Derivation of Eq. (\ref{EQ_MTH})}
\label{APP_C}

First note that for the input state $| \alpha_1 \rangle \dots |\alpha_M \rangle$ the interference operator reads
\begin{equation}
\hat{\mathcal{I}}_M = ( | \alpha_1 \rangle \langle \alpha_1 | - |0\rangle_1 \langle 0 |) \otimes \dots \otimes ( | \alpha_M \rangle \langle \alpha_M | - |0\rangle_1 \langle 0 |).
\label{EQ_IM_PROD}
\end{equation}
The higher-order interference term for the setup of Fig.~\ref{FIG_M} is
\begin{equation}
\begin{split}
\mathcal{I}_M = \mathrm{Tr} ( U_{1M}^\dagger \dots U_{13}^\dagger (a_1^\dagger a_1 + a_1^\dagger a_2 +\\
 a_2^\dagger a_1 + a_2^\dagger a_2) U_{13} \dots U_{1M} \hat{\mathcal{I}}_M ),
\end{split}
\end{equation}
where we have used the transformation of the beam splitter.
Note that the first and the last term in the inner bracket do not contribute to $\mathcal{I}_M$, because they commute with the unitaries and due to the partial trace property in Eq.~(\ref{EQ_PROP}).
From the definition of non-linear phase shift
\begin{equation}
\mathcal{I}_M = \mathrm{Tr} \left( a_1^\dagger a_2 \exp(-i \theta (a_3^\dagger a_3 + \dots + a_M^\dagger a_M)) \hat{\mathcal{I}}_M \right) + \textrm{c.c.}.
\end{equation}
Assuming all the input coherent states differ just by phases, i.e. $| \alpha_m \rangle = |\alpha e^{i \varphi_m} \rangle$, using $\exp(-i \theta a_m^\dagger a_m) | \alpha e^{i \varphi_m} \rangle = | \alpha e^{- i (\theta - \varphi_m)} \rangle$ and Eq.~(\ref{EQ_IM_PROD}) we find
\begin{equation}
\mathcal{I}_M = e^{i (\varphi_2 - \varphi_1)} \frac{1}{2} | \alpha|^2 \left( \langle \alpha | \alpha e^{- i \theta} \rangle - 1 \right)^{M-2} + \textrm{c.c}.
\end{equation}
Let us denote the complex coefficient multiplying the first exponential by $A = |A| e^{i \delta}$.
With this notation
\begin{eqnarray}
\mathcal{I}_M & = & 2 |A| \cos(\delta) \cos(\varphi_2 - \varphi_1) + 2 |A| \sin(\delta) \sin(\varphi_2 - \varphi_1) \nonumber \\
& = & 2 |A| \cos(\varphi_2 - \varphi_1 - \delta).
\end{eqnarray}
In Eq.~(\ref{EQ_MTH}) we additionally used the formula for the overlap between coherent states.

\subsection{Intensity under independent noise}
\label{APP_D}

Here we show that any detector noise independent of the input state shifts the measured intensity by a constant.
We model independent noise by adding an ancillary mode in the state $| d \rangle = \sum_n \sqrt{d(n)} | n \rangle$, 
and introduce measurement operators describing the detection of $n$ photons by the noisy detector as follows:
\begin{equation}
\Pi_n = \sum_{k = 0}^\infty | k \rangle \langle k | \otimes | n - k \rangle \langle n - k |,
\end{equation}
where the first Hilbert space describes the measured system, the second space is for the ancilla and all the kets where $n-k$ is negative are replaced by zeros.
Indeed one verifies that $r(n) = \mathrm{Tr}((\rho \otimes | d \rangle \langle d |) \Pi_n)$.
The intensity of this noisy measurement reads:
\begin{eqnarray}
\tilde{I} = \sum_{n = 0}^\infty n \, r(n) = \sum_{k = 0}^\infty \langle k | \rho | k \rangle  \sum_{n = k}^\infty n | \langle d| n-k \rangle |^2.
\end{eqnarray}
We write $n = (n-k) + k$ and accordingly split the second sum into:
\begin{eqnarray}
S_1 & = & \sum_{n = k}^\infty (n-k) | \langle d| n-k \rangle |^2 = \langle d^\dagger d \rangle, \\
S_2 & = & \sum_{n = k}^\infty k | \langle d| n-k \rangle |^2 = k,
\end{eqnarray}
where in the first line we introduced the number operator $d^\dagger d$ for the ancillary mode and in the second line we used the completeness relation.
The expectation value $\Delta = \langle d^\dagger d \rangle$ is calculated in the state $| d \rangle$ describing the noise.
Finally the noisy intensity is
\begin{eqnarray}
\tilde{I} = \sum_{k = 0}^\infty \langle k | \rho | k \rangle (\Delta + k) = \Delta + \langle a^\dagger a \rangle,
\end{eqnarray}
where $\langle a^\dagger a \rangle$ gives the intensity of the ideal measurement.

\input{hoi.bbl}

\end{document}

%% file: hoi.bbl
%